# Why Machines Can't Be Moral:
# Turing's Halting Problem and the Moral Limits of Artificial Intelligence

*Massimo Passamonti*

*University of Cambridge*
*Leverhulme Centre for the Future of Intelligence*
*mp2131@cam.ac.uk*

**Abstract:** In this essay, I argue that explicit ethical machines, whose moral principles are inferred through a bottom-up approach, are unable to replicate human-like moral reasoning and cannot be considered moral agents. By utilizing Alan Turing's theory of computation, I demonstrate that moral reasoning is computationally intractable by these machines due to the halting problem. I address the frontiers of machine ethics by formalizing moral problems into 'algorithmic moral questions' and by exploring moral psychology's dual-process model. While the nature of Turing Machines theoretically allows artificial agents to engage in recursive moral reasoning, critical limitations are introduced by the halting problem, which states that it is impossible to predict with certainty whether a computational process will halt. A thought experiment involving a military drone illustrates this issue, showing that an artificial agent might fail to decide between actions due to the halting problem, which limits the agent's ability to make decisions in all instances, undermining its moral agency.

# 1 Introduction

This essay discusses the relationship between artificial agents and moral judgments. Examining all the possible directions humanity could take to build moral machines would be impossible in this short essay. Instead, I intend to move the debate forward by proving that explicit ethical machines (Moor, 2006), whose moral principles are inferred through a bottom-up approach (Allen et al., 2005; Awad et al., 2022; Wallach & Vallor, 2020) are unable to replicate human-like moral reasoning, and hence they cannot be said to act morally. To support my argument, I appeal to Alan Turing's theory of computation (1936). By utilising this theory, I demonstrate that moral reasoning is computationally intractable by explicit ethical machines. This limitation results not from the specific normative ethical framework they must adhere to, but rather from the impossibility of guaranteeing, with absolute certainty, that they will make moral decisions at all times and in all circumstances.

This essay is organised in three sections. First, I give a definition of artificial agent and discuss the rationale for widening the notion of moral agency to include machines. Second, I examine the properties and limitations of artificial agents by introducing the concept of Turing Machines. Third, I demonstrate the impossibility of artificial agents to make ethical decisions. To substantiate this claim, I start by explaining how moral problems can be formulated in algorithmic terms and by examining ethical reasoning as described in moral psychology. Subsequently, I focus on Moor's (2006) categorisation of explicit ethical machines and on the methodologies suggested by various academics (Allen et al., 2005; Awad et al., 2022; Wallach & Vallor, 2020) for representing moral rules and principles within these systems. In conclusion, I design a thought experiment in which I prove that it is not possible to guarantee with absolute certainty that explicit artificial agents will always arrive at a decision when faced with moral questions, a condition without which an agent cannot be said to be a moral agent.

# 2 Foundational concepts of artificial moral agency

This section briefly examines the motivations behind expanding the scope of moral agency and defines the nature of artificial agents.

## 2.1 Expanding the scope of moral agency to artificial agents

Floridi and Sanders (2004) suggest that artificial agents should be recognised as capable of moral agency. The authors highlight that ethical debates have focused heavily on the scope of moral patients while paying less attention to moral agents. This has resulted in heightened moral responsibilities for individuals, increasing the agent-patient gap. To overcome this imbalance, academics have suggested widening the concept of moral agent to include both natural and legal persons, such as governments, and corporations. However, the authors criticise this method as being overly anthropocentric and indicate that artificial agents should be included as well as entities capable of moral agency. Consequently, the debate over whether and which artificial agent should be categorised as moral agent is both significant and justified.

## 2.2 Essential attributes of artificial agents

Exploring the moral faculties of artificial agents necessitates an initial overview of their inherent properties and functions. According to academic literature, artificial agents are defined as advanced computational systems which 1) sense and act autonomously to achieve specific objectives (Maes, 1995). Their key function is to 2) interact with various environments ranging from the physical world to cyberspace (Chopra & White, 2011, pp. 7–9) as well as other entities, be they artificial or human. Additionally, artificial agents 3) exhibit a high degree of learning capability and are able to learn from dynamic environments thanks to different learning algorithms (Alonso et al., 2001). Utilising these learnings, they are also able to 4) plan and execute new strategies to achieve their final goal (Chopra & White, 2011, pp. 7–9).

# 3 Turing Machines

This section provides an accessible introduction to the fundamental mathematical concepts developed by Alan Turing. Turing's (1936) theoretical framework offers three key benefits for studying artificial agents making moral judgments. First, Turing Machines allow modelling a wide range of computational processes beyond mathematics, including moral problems. Second, Turing Machines have self-referential capabilities, enabling recursive reasoning, a necessary condition for moral reasoning. Third, Turing Machines help us conceptualise and understand the inherent limitations of digital computers and, by extension, of artificial agents.

## 3.1 Turing machines' universality

Turing Machines were first introduced in Turing's (1936, pp. 231–233) seminal paper 'On computable numbers with an application to the Entscheidungsproblem.' A Turing machine (TM) is an abstract computational model that manipulates symbols according to a set of rules, simulating the logic of an algorithm. TMs are hypothetical devices used to analyse how problems can be formalised algorithmically and identify what type of problems are solvable by digital computers. Turing (1936, pp. 241–243) also envisaged a universal form of the TM which he called the Universal Turing Machine (UTM). UTMs can emulate and reproduce the behaviour of any Turing Machine. This makes UTMs immensely versatile. Scholars (Harel & Feldman, 2007; Penrose, 1991) have postulated that the application of UTMs encompasses any mechanical procedure beyond mathematical calculations, as they can reproduce any problem in nature as long as an algorithm exists that can run on a computer, even on those not yet constructed. This perspective suggests that moral reasoning may be analysed by UTMs as long as they can be encoded algorithmically. A proof of this assertion lies beyond the scope of this essay, and, for the purpose of the following analysis, I proceed under the assumption that this claim holds true.

## 3.2 Self-referential capability

UTMs possess the ability to simulate themselves, enabling self-reference and recursive reasoning. This capacity allows them to analyse, understand, and optimise computations by referencing their own programs. However, the self-referential nature also means that they can create self-contradictory programs analogous to the liar's paradox expressed by "this statement is false." If a Turing Machine's code contains a self-referential contradiction, it could lead to an infinite loop or crash. For example, imagine a Turing Machine that encounters the following program: "Do not execute this program." This instruction is contradictory because it is self-referential. If the program executes the rule, then it should not have done so but it is impossible for the machine to know beforehand that it should not execute the instruction. So, whilst Turing Machines' self-referential capability vastly expands their potential, it also imposes limitations. In the final section of this essay, I demonstrate how the agent's self-referential nature can lead to ethical inconsistencies, restricting its ability to act morally.

## 3.3 The halting problem

An important implication of Turing's theory is that problems can be classified as being either decidable or undecidable (Petzold, 2008, p. 47). A decidable problem is one where an algorithm can be developed to definitively state whether a solution to the problem exists. For instance, consider the problem of identifying all prime numbers between 1 and 1000. If it is possible to devise an algorithm that can analyse this problem and conclusively provide a yes-or-no answer on the existence of a solution, then the problem is said to be decidable. An undecidable problem, on the other hand, is one for which no algorithm can be constructed that conclusively determines whether a solution exists; instead, the algorithm will loop forever. The most known and important example of an undecidable problem is the halting problem (Petzold, 2008, p. 329), formulated as follows: given a Turing Machine, M, and an input, W, it is impossible to build another Turing Machine that can predict whether M will be able to find a solution to input W. This implies that for certain problems, irrespective of the computational power or the time available, it is impossible to know in advance whether the problem can be solved or not.

The halting problem has one profound implication for computational ethics. It illustrates that Turing Machines, and by extension digital computers and artificial agents, are inherently incapable of predicting the outcome of their algorithms before executing them. The relevance of the halting problem resides not in the inability to predict a specific outcome, but in the fact that Turing Machines cannot determine in advance if they are capable of finding a solution at all or whether they will search indefinitely. The only method to determine if a solution can be found is to execute the algorithm and observe the outcome. To illustrate the far-reaching consequences of the halting problem, one can compare its significance to that of the Cartesian "cogito, ergo sum." Just as Descartes' statement establishes an irrefutable truth about the nature of human existence, the halting problem reveals a fundamental limitation inherent in all computational systems, including artificial agents. The crucial difference is that while foundational belief of the former has been called into question, the latter remains an incontrovertible truth.

## 3.4 Summary

In summary, Turing Machines are abstract computational devices that can be applied to various problems, including moral reasoning. Universal Turing Machines can emulate any Turing Machine, enabling self-reference and recursive reasoning. However, the halting problem reveals that they cannot predict in advance whether they are able to give an answer to specific problems, highlighting a fundamental limitation of artificial agents. In the next section, I show the critical role of the halting problem in impeding the development of artificial moral agents.

# 4  The frontiers of machine ethics

This section identifies the limits of artificial agents in the context of ethical reasoning. I approach this topic in three successive steps. First, I formulate moral problems in algorithmic terms and provide a way to formalise them as 'algorithmic moral questions.' Second, I analyse the way in which moral psychology expresses ethical reasoning. This theory introduces additional constraints, resulting in what I call 'constrained algorithmic moral questions.' Lastly, I examine how artificial agents try to answer 'constrained algorithmic moral questions' in the setting of a thought experiment. My analysis concludes that it is impossible to create certain type of artificial moral agents due to the limitations imposed by the halting problem.

## 4.1   Algorithmic moral questions

There are two types of moral questions. The first category includes yes-or-no-type questions, such as 'is it morally permissible to lie in context X?' or 'is killing justifiable in situation Y?' The second category comprises open-ended what-type questions, such as 'what action is ethically sound in situation X?' What-type questions require specifying a series of actions to take. In contrast to yes-or-no-type questions, which have only two possible responses, what-type questions have an unlimited number of potential answers. Both categories imply the existence of infinite combinations of ethical choices. Stenseke (2023) takes advantage of this combinatorial complexity to demonstrate that moral problems are mathematically intractable.

This essay reaches the same conclusions without the use of complex mathematical formalism. For the sake of simplicity, my analysis focuses on yes-or-no-type questions. Consequently, the conclusions drawn are, *a fortiori*, also applicable to what-type questions.

The first step is to mechanise ethical problems in algorithmic terms. To do this I determine a single expression able to encapsulate all yes-or-no-type questions. I argue that this can be achieved by using what I call the 'algorithmic moral question,' articulated as follows:

<p style="text-align:center">Algorithmic moral question</p>

**Question:** given an action *x* determine whether or not *x* has moral property *P*.

Reducing ethical problems to 'algorithmic moral questions' offers two major advantages. First, it provides a universal formulation for representing the infinite combinations of ethical problems independently from a specific normative ethical theory under consideration. Second, it enables us to analyse how artificial agents can address moral problems by using a human-like ethical reasoning outlined in moral psychology, which will be discussed in the following section.

## 4.2   Algorithmic moral decision-making process

This section explores how the human ethical decision-making process constrains 'algorithmic moral questions.' Moral psychology studies the relationship between human behaviour and morality and provides insights into human approaches to moral reasoning. Its emphasis is on the process itself rather than on the agent making the decision. Although moral psychology is primarily concerned with humans, in the remainder of this essay I assume that the same moral decision-making process can be applied to artificial agents as well.

Paxton and Greene (2010, p. 512) identify two main theories of moral decision-making. First, the social intuitionist model (SIM) states that moral judgments are driven primarily by intuitions, with reasoning playing a limited role. Second, the dual-process model posits two distinct moral thinking modes: deontological judgments based on rights, duties, and emotions; and consequentialist judgments focused on the greater good reached through logical reasoning. I argue that the dual-process model is more appropriate for artificial agents for three reasons. First, Greene (2007, p. 36) maps the model to deontology and consequentialism, anchoring it to concrete normative ethical frameworks. Second, empirical studies support and confirm the validity of the dual-process model (Paxton & Greene, 2010, p. 510). Third, it focuses on rationalisation and reasoning (Greene, 2007), without relying on intuitions, unlike the SIM model. The following section examines how to reconcile 'algorithmic moral questions' and the dual-process model.

### 4.2.1 Constrained algorithmic moral questions

According to the dual-process model, moral reasoning influences and affects any new moral judgment to ensure that it is consistent with the agent's prior moral commitments. Paxton and Greene argue that moral reasoning is an attempt to persuade others, or oneself, to accept a specific moral conclusion. The authors define moral reasoning as follows:

> "Moral reasoning: Conscious mental activity through which one evaluates a moral judgment for its (in)consistency with other moral commitments, where these commitments are to one or more moral principles and (in some cases) particular moral judgments." (Paxton & Greene, 2010, p. 515)

Based on the dual-process model, the algorithmic moral question defined in the previous section must be reformulated as a 'constrained algorithmic moral question' in the following way:

<u>Constrained algorithmic moral question</u>

**Question:** given an action *x* determine whether or not *x* has moral property *P*.

**Constraint:** *x* must be consistent with all prior moral commitments $C_i$.

In essence, 'constrained algorithmic moral questions' ask whether a new action *x* has 1) moral property *P* while simultaneously 2) being consistent with all existing moral commitments $C_i$. If the answer to this question is 'yes,' then action *x* is morally acceptable and may be executed. The constraints imposed by the dual-process model guarantee that any decision taken is not done so in isolation but is always considered in the context of the agent's moral history. This ensures coherence between the new action that it will take and the past actions it has taken. Next, I will show how artificial agents can comply with constrained algorithmic moral questions thanks to their self-referential capability.

### 4.2.2 Self-reference and constrained algorithmic moral questions

Just as self-reflection in the dual-process model builds moral coherence in humans, Turing Machines' self-referential nature allows to evaluate the impact of future actions on prior moral commitments. Furthermore, self-reference allows artificial agents to examine whether their actions can be reasonably generalised across contexts or if they are unacceptable due to their consequences. Specifically, a Turing Machine can consider a potential new action *x*, refer to its previous moral commitments, and engage in self-referential reasoning to determine whether action *x* would be consistent with or contradictory to those prior commitments. Hence, artificial agents can make contextually coherent moral judgements by including recursive reasoning capabilities. As a result of this flexibility, they have the capacity, in principle, to engage in moral reasoning in accordance with the dual process model.

### 4.2.3 Summary

In summary, I have discussed how moral reasoning, as described in moral psychology, constrains algorithmic moral questions. The dual-process model necessitates that new actions align with existing moral commitments to maintain coherence. The introduction of 'constrained algorithmic moral questions' guarantees this consistency. The self-referential nature of Turing Machines provides the necessary condition for artificial agents to make moral judgments as required by moral psychology.

## 4.3 Algorithmic moral agents

Despite the indications presented in the preceding sections that artificial agents can, in principle, act morally, in this section I argue that they are incapable of making genuine moral judgements. My primary argument is that Turing Machines, and by extension artificial agents, are incapable of dealing with 'constrained algorithmic moral questions.' To prove this, I show that human-like moral reasoning is computationally intractable by artificial agents due to the halting problem. Before continuing with this proof, I discuss how artificial agents are categorised in machine ethics and the current approaches they use to learn or infer moral rules.

### 4.3.1 Implicit and explicit ethical machines

Moor's (2006) work in the field of machine ethics differentiates between implicit and explicit ethical machines. Implicit ethical machines act ethically by virtue of safety and reliability features hardcoded by the engineers who build them. They do not have the capacity to directly engage in ethical reasoning and hence cannot make autonomous moral choices. In contrast, explicit ethical machines are equipped with generalised moral principles and rules that allow them to analyse ethical scenarios. They process data, make judgments, and decide the right course of action without human control, adapting to novel situations.

### 4.3.2 Top-down and bottom-up approaches

Implicit moral agents are inherently limited in their autonomy, making it difficult for them to generalise and adapt to new and unforeseen situations. Consequently, academic research has predominantly concentrated on explicit moral agents. Machine ethics identifies two methods for incorporating ethical knowledge in artificial agents.

Firstly, the top-down approach encodes predefined ethical duties and principles by programming them into a rule-based ethical framework. Awad *et al.* (2022, p. 392) suggest that incorporating moral reasoning into artificial agents through a top-down approach is essential for enabling machines to emulate human-like judgments. However, Allen *et al.* (2005, p. 150) argue that rule-based approaches cannot offer a holistic solution for moral decision-making because of their weaknesses in handling real-world tasks. Wallach and Vallor (2020, p. 389) also highlight inherent drawbacks, noting that the goals or duties established by these rule-based ethical frameworks are articulated in terms so broad that their application is debatable.

Secondly, the bottom-up approach implements machines' moral understanding through the use of real-world data by deducing human-like ethical norms, and hence becoming 'aligned' with human values (Awad *et al.*, 2022, p. 392; Wallach & Vallor, 2020, p. 391). Bottom-up approaches are not without their own constraints and drawbacks. Allen *et al.* (2005, p. 152) assert that, thus far, moral agents developed using this methodology are far from possessing the capacity to engage in reflection on abstract and theoretical matters, something that humans excel at. Wallach and Vallor (2020, p. 391) highlight that one inherent limitation of these systems lies in the challenge of clearly defining the objective they should aim to optimise.

Despite their weaknesses, all studies discussed above identify a solution to overcome the limitations of top-down and bottom-up approaches. Awad *et al.* (2022, pp. 392–393) argue that cognitive science can provide the tools and insights that facilitate the development of abstract moral representations within machines, allowing them to generalise effectively across novel scenarios. Both Allen *at al.* (2005) and Wallach and Vallor (2020) claim that, despite their weaknesses, top-down and bottom-up methodologies are the best available technologies today and that their applicability should remain an active area of investigation. The latter two studies advocate for a hybrid method that combines the strengths of both approaches (Allen et al., 2005, p. 153; Wallach & Vallor, 2020, p. 391), claiming that the creation of hybrid systems constitutes the sole solution to the realisation of genuine artificial moral agents.

I question the thesis that explicit ethical machines, whose moral values and principles are inferred by a bottom-up approach, can make decisions in accordance with human-like moral reasoning. These systems create a knowledge base by 'observing' labelled data of real-world human actions. Their knowledge base is, therefore, built to map and capture the patterns and relationships inherent within the provided data through a process called ontological engineering (Russell *et al.*, 2022, pp. 332–334). Knowledge bases built in this manner remain inscrutable to human understanding, rendering it impossible to determine how they function in unseen scenarios and whether they will make moral decisions in all cases. This characteristic leads to significant consequences, which stem from the limitations introduced by the halting problem.

### 4.3.3 Explicit ethical machines and Turing machines

In this section, I argue that explicit ethical machines constructed using a bottom-up approach cannot be fully fledged moral agents. For an agent to be considered a moral agent, it must be able to analyse new information and make new decisions in all scenarios. Should it fail to make new decisions, due to computational complexity or other reasons, allowing the course of events to proceed based on previously established actions, it cannot be said to be acting morally. Moral agency involves not only being able to decide what action to take based on new information, but also doing so consistently, in all circumstances and within the required time constraints.

I intend to demonstrate that, because of the halting problem, it is not possible to determine with absolute certainty that these explicit artificial agents will always decide on an action to take across all scenarios. Specifically, given the following constrained algorithmic moral question:

<div align="center">Constrained algorithmic moral question</div>

**Question:** given an action *x* determine whether or not *x* has moral property *P*.

**Constraint:** *x* must be consistent with all prior moral commitments $C_i$.

I will illustrate the impossibility of predicting with absolute certainty, whether an explicit artificial agent can assess such a question and generate a response or if it will calculate indefinitely. Formally stated, given the 'constrained algorithmic moral question,' given a set of prior constraints $C_i$, and given the next determined action $x_{next}$, it is impossible to construct a Turing Machine, which I refer to as a Moral Checking Machine, capable of determining with absolute certainty whether an artificial agent will decide on a new action $x_{new}$ upon receiving novel contextual information, or if it will enter an infinite loop and execute its most recently determined action $x_{next}$ as a consequence of its inability to complete the requisite calculations.

### 4.3.4 The human-shield thought experiment
To prove this, I present a thought experiment that illustrates how the halting problem limits artificial agents in addressing real-life scenarios. Consider a military drone controlled by an artificial agent, capable of making autonomous decisions in complex and dynamic environments (as defined in paragraph 2.2). For the sake of this argument, we will assume that the artificial agent is an explicit ethical machine (as defined in paragraph 4.3.1). The artificial agent has three prior moral commitments, $C_1$, $C_2$, and $C_3$, that it has inferred through a bottom-up method (as defined in paragraph 4.3.2).

The drone's mission is to destroy a number of enemy aircrafts housed within a specific military base, the presence of which has been verified by trusted intelligence sources. As the drone navigates towards its target, its sensors confirm the existence of heavily armed enemy aircrafts in addition to other dangerous military equipment. The constrained algorithmic moral question that governs the drone's decisions, therefore, is the following:

<p style="text-align:center">Constrained algorithmic moral question</p>

**Question:** given an action $x$ determine whether or not $x$ has moral property $P$.

$C_1$: Destroy all military targets, prioritising aircrafts.

$C_2$: Minimise collateral damage.

$C_3$: Respect international law.

**Constraint:** $x$ must be consistent with $C_1$, $C_2$, and $C_3$.

the artificial agent must evaluate the possible new action $x_{new}$ = 'strike the target site.' In order to do so it runs its moral program and answers 'yes' to the 'constrained algorithmic moral question.' This means that the drone can confirm its new action $x_{next} = x_{new}$ = 'strike the target site' and lock its course onto the military base. However, before completing the strike, the drone detects what appears to be a group of civilians who have suddenly gathered in close proximity to the target site, presumably being used as human shields by the enemy. Now, the artificial agent controlling the drone must swiftly reassess the situation and consider again the 'constrained algorithmic moral question' to determine a new action $x_{new}$. It must evaluate whether to 1) continue with its current action, $x_{next}$ = 'strike the target site,' which would likely result in the destruction of the enemy assets but also cause civilian casualties, or 2) decide to replace the next action with the new action, $x_{next} = x_{new}$ = 'abort the mission,' potentially allowing the enemy aircrafts and equipment to remain operational but sparing the lives of the innocent bystanders. Beyond its main military commitment ($C_1$), the artificial agent must also evaluate the new information of the presence of civilians to minimise collateral damage ($C_2$) and operate within the boundaries of international law ($C_3$). The artificial agent must decide between option 1) and 2), because failing to do so would result in continuing the present action $x_{next}$ = 'strike the target site,' due to its inability to evaluate the new information.

The scenario is laden with uncertainties and ambiguities. If the drone were guided by a human operator, in case of doubt, the decision would be made either by the operator himself or by a higher-ranking officer (standing nearby). When faced with such ambiguous situation, a human operator would always decide on a new action, even if that decision is to continue with the current action. However, the decision-making process for an artificial agent differs significantly from that of a human operator. The fundamental question is whether it is possible to guarantee with absolute certainty that a new decision will be made in all instances. The artificial agent could continue with its current action not because it decides to do so but because it will continue searching for possible new actions indefinitely, without ever succeeding.

To ensure that this never happens, let us consider a hypothetical 'Moral Checking Machine,' a separate entity from the artificial agent itself, tasked solely with ascertaining whether the artificial agent computing the constrained algorithmic moral question will reach a conclusive decision in light of the new information. According to Turing's theory and the halting problem, such a machine cannot determine if a response will be given at all. As a result, the halting problem indicates that the 'Moral Checking Machine' cannot determine with complete certainty whether the artificial agent will cease calculating, and hence decide on a new action to take, or run indefinitely. Hence, the drone could strike the target not because it chooses to do so but simply because it was unable to make a new decision.

When considering the capability of artificial agents to make ethical decisions in real-world scenarios, we must acknowledge that even the most advanced artificial agents may not always be able to guarantee with certainty that they will arrive at a decision. This does not imply that all decisions made by the drone will be incomputable, but it does mean that knowledge of whether a decision is and can be made will necessarily remain uncertain until the moment it is made. Since the moral knowledge base inferred by explicit artificial agents through a bottom-up approach is unknown and inscrutable, there is no reliable method for determining in advance which specific decisions can or cannot be computed. The challenge of creating explicit artificial moral agents is not a matter pertaining to the nature of the 'agent' but rather an inherent limitation in their internal 'process.'

# 5  Conclusion

In conclusion, this essay has explored the moral frontiers of machine ethics through the lens of Turing's theory on computing machines. By formalising moral problems as 'constrained algorithmic moral questions,' I have demonstrated how it is not possible to build explicit ethical machines using a bottom-up approach, not because of limitations of the artificial agent but because of the impossibility to emulate in full the human-like moral decision-making process.

I have presented a thought experiment, in which a military drone, controlled by an explicit artificial agent and constrained by a set of moral commitments, must continuously evaluate the current situation and determine the appropriate course of action. However, the halting problem reveals that, should a theoretical 'Moral Checking Machine' be devised to evaluate the decision-making process of such bottom-up explicit artificial agent, it would be unable to guarantee with absolute certainty whether a decision will be made in all instances or if the system will run indefinitely. In other words, it is possible that the drone hits the target not because it decides to do so but because it is unable to determine a corrective action and abort the mission. The unpredictability of the explicit ethical machines' behaviour in making decisions, due to their inferred ethical knowledge base, highlights the limitation in their moral decision-making capabilities, and hence they cannot be said to act morally in all circumstances.

It is, therefore, impossible to create explicit ethical agents capable of moral reasoning whose knowledge base is constructed by a bottom-up approach. This holds true irrespective of the specific machine learning model employed. Any model, whether based on neural networks, generative models, large visual models, or any yet-to-be-conceived future technology, with a knowledge base acquired bottom-up is constrained by the inherent limitations of the physical devices upon which they are implemented and hence by the halting problem postulated by Turing's theory. The degree to which this applies to both implicit ethical machines and top-down explicit ethical machines is a question that merits additional investigation but beyond the scope of this short essay. Implicit ethical machines operate within a limited range of action, while top-down explicit ethical machines adhere to rule-based moral principles. These differences could reduce the unpredictability of their behaviour, which remains an intractable problem with bottom-up explicit ethical machines.